# Kernel-based Detection of Coincidentally Correct Test Cases to Improve Fault Localization Effectiveness

Farid Feyzi*, Saeed Parsa

Farid_feyzi@comp.iust.ac.ir, Parsa@iust.ac.ir

School of Computer Engineering, Iran University of Science and Technology, Tehran, Iran

**Farid Feyzi** received the M.S. degree in Software Engineering from the Sharif University of Technology in 2012. He is currently a Ph.D. Candidate in the Department of Computer Engineering at Iran University of Science and Technology. His research focus is on developing statistical algorithms to improve software quality with an emphasis on statistical fault localization and automated test data generation.

**Saeed Parsa** received his B.Sc. in Mathematics and Computer Science from Sharif University of Technology, Iran, his M.Sc. degree in Computer Science from the University of Salford in England, and his Ph.D. in Computer Science from the University of Salford, England. He is an associate professor of Computer Science at the Iran University of Science and Technology. His research interests include software engineering, software testing and debugging and algorithms.

## Abstract

Although empirical studies have confirmed the effectiveness of spectrum based fault localization (SBFL) techniques, their performance may be degraded due to presence of some undesired circumstances such as the existence of coincidental correctness (CC) where one or more passing test cases exercise a faulty statement and thus causing some confusion to decide whether the underlying exercised statement is faulty or not. This article aims at improving SBFL effectiveness by mitigating the effect of CC test cases. In this regard, a new method is proposed that uses a support vector machine (SVM) with a customized kernel function. To build the kernel function, we applied a new sequence-matching algorithm that measures the similarities between passing and failing executions. We conducted some experiments to assess the proposed method. The results show that our method can effectively improve the performance of SBFL techniques.

## 1 Introduction

The process of locating the bug and identifying its cause is referred to as software fault localization, which is one of the most labor-intensive activities of debugging [1]. Due to intricacy and inaccuracy of manual fault localization, a considerable amount of research has been carried out to develop automated techniques and tools to assist developers in finding bugs [1-10]. A class of these methods referred to as SBFL techniques estimate the suspiciousness of each program statement by analyzing the spectra of passing and failing test cases. They attempt to rank program elements according to their presence (as well as absence) in failing and passing executions. The more correlation between such program elements and the presence of the observed failures, the larger degree of suspiciousness is assigned to the element.

Recently, Masri et al. [11] conducted an empirical study that reveals most dynamic information flows in programs do not convey any measurable information, which means that it is likely that many infectious states might not propagate to the output, thus, leading to coincidental correctness [11-14]. In this case, the defect is reached but either the program is not transitioned into an infectious state, or the infection is not propagated to the output. Masri et al. and several other researchers, demonstrated that coincidental correctness is prevalent and is a safety reducing factor for SBFL [11-17]. That is, when coincidentally correct tests are present, the faulty statement will likely be ranked as less suspicious than when they are not present.

As stated in [20], the problem of identification of test cases with CC property can be considered as an instance of the mislabeled data identification in machine learning. In the scenario, we may regard CC test cases as mislabeled data, i.e. test cases whose test status should be changed to "failed". Therefore, existing machine learning techniques in detecting mislabeled data can be easily adapted to address this problem. Kernel methods are currently at the heart of many machine learning algorithms [29]. These methods attempt to project the data into a higher dimensional embedding, known as the Hilbert space and then search for linear relations in the transformed space.

In this article, we present a novel kernel-based approach for detecting CC runs in a test suite. Unlike other approaches, we directly formulate CC runs detection as a constrained optimization problem and introduce a kernel-based strategy to find them. In this regard, a new sequence-matching algorithm [21] is applied to find the common non-overlapping sub-paths among different program execution paths. This algorithm could be used to build a kernel function for a support vector machine (SVM) classifier. The SVM classifier constructs a hyperplane which separates failing and passing training execution vectors by a clear gap that is as wide as possible in the program feature space. The trained SVM classifier could be then used to address the identification of CC test cases by comparing program execution paths with the hyperplane [22][23].

The remainder of this article is organized as follows. Section 2 presents a motivating example that shows how CC has a safety reducing effect on SBFL. The details of the proposed method are described in section 3. The experiments and results are shown in section 4. The related works and threats to validity are presented in sections 5 and 6. Finally, the concluding remarks are mentioned in section 7.

## 2 Motivating Example

To illustrate the impact of the existence of CC on suspiciousness scores, we introduce a simple example. In the following, we use Tarantula [10], Ochiai [8] and Naish [9] to measure the suspiciousness metric of predicates.

Tarantula  $$susp(p) = \frac{N_{CF}(p)/N_F}{N_{CS}(p)/N_S + N_{CF}(p)/N_F} \tag{4}$$

Ochiai  $$susp(p) = \frac{N_{CF}(p)}{\sqrt{N_F \times (N_{CS}(p) + N_{CF}(p))}} \tag{5}$$

Naish  $$susp(p) = \begin{cases} -1 & if\ N_{UF}(p) > 0 \\ N_{US}(p) & otherwise \end{cases} \tag{6}$$

In Equations (4-6), p is a predicate; $N_{CF}(p)$ is the number of failed tests that execute p; $N_{UF}(p)$ is the number of failed tests that do not execute p; $N_{CS}(p)$ is the number of passed tests that execute p; and $N_{US}(p)$ is the number of passed tests that do not execute p. As shown in Table 1, there are twelve tests t1- t12, in which t1,t2, … t6 are failed, t7, t8, …, t12 are passed, and t7, t8 and t10 are cc. There are five predicates p1; p2; p3; p4; p5, in which p2 is the faulty predicate. The coverage information is shown in Table 1. The Tarantula, Ochiai and Naish values of predicates are calculated by Equations (4-6). It is not difficult to see that p1 the most suspicious one and the faulty one p2 is masked. If we can identify that t7, t8 and t10 are CC and flip their outcome states from pass to fail, then suspiciousness values are changed to suspiciousness*, represented in Table 2. The results show that the presence of CC test cases confuses fault localization computations in accurately estimating the suspiciousness of statements.

Table 1. An example including a calculator source code and predicates coverage information.

| # | Program | | Input Parameters | | | Predicates | | | | | TC outcome state | |
|---|---|---|---|---|---|---|---|---|---|---|---|---|
| | | | a | b | c | P1 | P2 | P3 | P4 | P5 | #tc | Fail/Pass |
| 1 | int Sample (int a, int b, int c){ | | | | | | | | | | | |
| 2 | int rsum, rdiv, result, rlog=0; | | 4 | 1 | 3 | 1 | 1 | 0 | 0 | 1 | t1 | Fail |
| 3 | result = 0; | | 7 | 6 | 3 | 1 | 1 | 0 | 0 | 1 | t2 | Fail |
| 4 | rdiv = 1; | | 6 | 3 | 3 | 1 | 1 | 0 | 0 | 1 | t3 | Fail |
| 5 | rsum = a + b; | | 2 | 1 | 3 | 1 | 1 | 0 | 0 | 1 | t4 | Fail |
| 6 | if ((a > 0) && (b > 0)) | | | | | | | | | | | |
| 7 | rdiv = a / b; | //P1 | 3 | 2 | 3 | 1 | 1 | 0 | 0 | 1 | t5 | Fail |
| 8 | rmax = b; | | 9 | 7 | 3 | 1 | 1 | 0 | 0 | 1 | t6 | Fail |
| 9 | if (a > b) | | | | | | | | | | | |
| 10 | rmax = b; //Bug; | //P2 | 8 | -3 | 2 | 0 | 1 | 0 | 1 | 0 | t7 | Pass (CC) |
| 11 | if (c = = 1) | | 9 | -2 | 2 | 0 | 1 | 0 | 1 | 0 | t8 | Pass(CC) |
| 12 | result = rsum; | //P3 | -6 | 8 | 3 | 0 | 0 | 0 | 0 | 1 | t9 | Pass |
| 13 | if (c = = 2) | | | | | | | | | | | |
| 14 | result = rdiv; | //P4 | 7 | 6 | 1 | 1 | 1 | 1 | 0 | 1 | t10 | Pass(CC) |
| 15 | if (c = = 3) | | | | | | | | | | | |
| 16 | result = rmax; | //P5 | 6 | 8 | 3 | 1 | 0 | 0 | 0 | 1 | t11 | Pass |
| 17 | return result} | | -8 | 9 | 3 | 0 | 0 | 0 | 0 | 1 | t12 | Pass |

Table 2. The negative impact of coincidental correctness.

| Localization Method | Suspiciousness | | | | | Suspiciousness* | | | | |
|---|---|---|---|---|---|---|---|---|---|---|
| | **P1** | P2 | P3 | P4 | P5 | P1 | **P2** | P3 | P4 | P5 |
| Tarantula | **0.75** | 0.67 | 0 | 0 | 0.67 | 0.875 | **1** | 1 | 1 | 0.67 |
| Ochiai | **0.86** | 0.81 | 0 | 0 | 0.81 | 1.01 | **1.22** | 0.4 | 0.58 | 0.81 |
| Naish | **4** | 3 | 1 | 1 | 3 | 1 | **3** | 1 | 1 | 1 |

## 3 Methodology Overview

The proposed SVM-based approach mainly consists of three steps: data collection, model training, and model construction. These steps are explained in detail in the following sub-sections.

### 3.1 Data collection

To collect the program execution data, programs are instrumented by inserting probes before their predicates. In our work, predicates are designed to capture the results of conditional statements and function calls [24][25]. We have instrumented return statements by inserting six predicates $C \leq 0, C < 0, C > 0, C \geq 0, C = 0$ and $C != 0$ for each return value C. During an execution of the instrumented program, execution paths are kept in a distinct vector, called execution vector. The program termination state is kept in the last cell of the vector.

### 3.2 Model training

At training phase, a binary classifier model is built. Then, a set of testing samples is given to the classifier to decide which classes the samples belong to. In the proposed approach, a binary classifier, $M$, is built for program execution paths to accurately predict whether an execution path is a pass or a failure. The classifier $M$ can be built based on a training set $\{(e_1, ce_1), \dots, (e_n, ce_n)\}$ where $\{e_1, \dots, e_n\}$ indicates the set of execution paths and the class labels $ce_1, \dots, ce_n$ can be either -1 or 1. The class labels -1 and 1 represent failing and passing executions, respectively. Since the execution profiles of CC test cases are expected to be similar to failing test cases [18-20], the resultant SVM classifier can be used to accurately predict whether a passing test execution is a true passing or a CC.

A major difficulty with applying linear classifiers is the generalization error caused by the close similarities amongst program executions [21]. If program executions are not linearly separable, a feature expansion technique [22] could be used to map the executions into a feature space in which the mapped executions will be linearly separable. SVM is a well-known classifier which applies feature expansion techniques to distinguish data items [23]. After using feature expansion technique, SVM classifier builds a hyperplane to separate the mapped executions into two classes with maximum possible margin within the feature space. Each execution, lying on the hyperplane is called a support vector.

To determine whether an execution path is a failure, the inner product of its corresponding vector with all the support vectors is computed.

### 3.3 Model construction

Suppose there is a set of $N$ predicates $P = \{p_1, p_2, \ldots, p_N\}$ in a program $R$. For program $R$, there are execution vectors $V = \{v_1, v_2, \ldots, v_m\}$, where each execution vector $v_i$ in $V$ corresponds to an execution path $E_i$, in the program execution space. Each execution path is classified as failing or passing according to the program termination state. The two classes are represented as members of set $C = \{1, -1\}$, where 1 and -1 indicate passing and failing, respectively. To build a classifier, an optimal weight vector $W$ could be built that determines whether an execution vector $v_i$ belongs to failing or passing partition of the program execution space. In fact, $W$ represents a vector perpendicular to hyperplane $H$, dividing the execution paths into two regions of failing and passing [21]. To determine whether a passing test case $p_i$ with corresponding execution path $E_i$, should be considered as fail or pass, the inner product of $v_i$ and $W$ could be computed as shown in relations (7) and (8).

$$W \cdot v_i - b \geq 1 \rightarrow e_i \text{ leads to pass} \rightarrow C_{e_i} = 1, p_i \text{ will be classified as a true passing test} \tag{7}$$

$$W \cdot v_i - b \leq 1 \rightarrow e_i \text{ leads to fail} \rightarrow C_{e_i} = -1, p_i \text{ will be classified as a CC test} \tag{8}$$

The parameter $b$ in relations (4) and (5) indicates the distance of hyperplane $H$, from the origin of the program execution space. The distance between two regions, separated by $H$, is maximized when the norm of the weight vector $W$ is minimized [21-22]. Therefore, the problem of finding the best hyperplane could be defined as the following optimization problem

$$\text{minimize } \tfrac{1}{2} ||W||^2 \text{ subject to } (C_{e_j} \cdot W \cdot v_j - b) \geq 1 \tag{9}$$

The optimization problem in relation (9) could be converted into a Lagrange function as follows

$$\text{minimize } L(W, b, \alpha) = \tfrac{1}{2} ||W||^2 - \sum_{j=1}^{m} \alpha_j \left( C_{e_j} \cdot W \cdot v_j - b \right) \text{ on the region } \alpha > 0 \tag{10}$$

After differentiating $L$ with respect to $W$ and $b$, we obtain the following relation

$$W = \sum_{j=1}^{m} \alpha_j \cdot C_{e_j} \cdot v_j \tag{11}$$

Now, substituting $W$ in relation (10) yields the following dual problem with respect to the Lagrange variable $\alpha$

$$\text{maximize } \left( \sum_{j=1}^{m} \alpha_j - 0.5 \sum_{j=1}^{m} \sum_{k=1}^{m} \alpha_j \alpha_k (v_j \cdot v_k) C_{e_j} \cdot C_{e_k} \right) \text{ on the region } \alpha > 0 \tag{12}$$

The inner product $(v_j \cdot v_k)$, in relation (9), represents the degree of similarity between two execution paths $e_j$ and $e_k$. Since program executions might be very similar and not linearly separable, it is required to map the original executions into a new feature space, in which the newly mapped executions could be linearly separable [28]. Linear separation is achieved by increasing the dimensions of the program execution space with combinations of its existing dimensions. To achieve this, feature expansion techniques could be applied by SVM. The SVM kernel function $K$, is capable of measuring the similarity between the converted vectors without direct mapping. Applying $K$ function, problem of finding the most suitable hyperplane $H$, could be expressed as follows:

$$\text{maximize } \left( \sum_{j=1}^{m} \alpha_j - 0.5 \sum_{j=1}^{m} \sum_{k=1}^{m} \alpha_j \alpha_k K(v_j \cdot v_k) C_{e_j} \cdot C_{e_k} \right) \text{ on the region } \alpha > 0 \tag{13}$$

A few number of general kernels such as polynomial, RBF, and Gaussian RBF kernels are available for applying in SVM [21-22]. A major difficulty with applying such predefined kernels is that they do not take any consideration for the sequence of predicates execution when building new dimensions for the feature space. As we claimed in [21], the execution order of program predicates is one of the primary parameters in discriminating failing from passing executions. To consider this fact, the SVM kernel is provided with a sequence matching function which is described in the next section.

### 3.4 Sequence-matching algorithm

In this section, a similarity assessment algorithm is presented to find all non-overlapping common subsequences between any two given sequences $s, t$. The algorithm uses the length and the number of common subsequences to measure the similarity between $s$ and $t$ [21]. The algorithm contains four major steps. In the first step a match matrix, **M** is constructed. Each element $M_{s,t}[i,j]$ is set to 1 if the $i$th predicate in $s$ is equal to the $j$th predicate in $t$. In the second step, the algorithm finds sub-diagonals of the continuous 1's in **M**. All the detected sub-diagonals are kept in a set, 'diagonal-set'. In the third step, the longest sub-diagonal of 1's in 'diagonal-set' is chosen and marked as selected. After a sub-diagonal is selected, to avoid overlaps between selected sub-diagonals, all the 1's in the columns and rows overlapping with the selected sub-diagonal are replaced with 0 and their corresponding sub-diagonals in 'diagonal-set' are either modified or totally removed. This process of selecting the longest non-overlapping sub-diagonals in 'diagonal-set' is repeated as long as the set includes unmarked sub-diagonals. Finally, the degree of similarity between $s$ and $t$ is measured as follows

$$sim(s,t) = \frac{\sum_{i=1}^{k} |match_i|}{k \times \max(|s|,|t|)} \tag{14}$$

In relation (14), $k$ indicates the number of matched subsequences and $match_i$ represents the $i$th longest matched subsequence between $s$ and $t$. As an example consider two sequences $s = \langle p_1, p_2, p_2, p_3, p_4, p_5, p_6 \rangle$ and $s = \langle p_1, p_2, p_3, p_4, p_5, p_7, p_5 \rangle$. Applying relation (14), the similarity between the predicate sequences $s$ and $t$ is computed as follows

$$\frac{1 \times 1 + 1 \times 4}{2 \times \max(7,7)} = \frac{5}{14}$$

**Algorithm** Maximal *Non-Overlapping Sequence Matching*

**Input**: Sequences $s$ and $t$
**Output**: Degree of similarity between $s$ and $t$.

**1)** Build a $match$ matrix for the sequences $s$ and $t$. $M_{s,t} = \begin{cases} 1 & if\ s[i] == t[j] \\ 0 & otherwise \end{cases}$

**2)** Compute diag-set=all continuous 1's in all diagonals of $match$ matrix
    **2-1)** Look for continuous 1's in diagonals above the main diagonal
    *len*=0;
    **for** (i=1 to |t|) {                          // for all the elements of the sequence $t$
      k=i;                                   //k indicates the column number.
      *flag*=False;
      **for** (j=1 to |s|){                    //find continuous 1's on the $i$th diagonal
        **if** ($M_{s,t}$[j,k]==1){
          **if** (!flag)  first=(j,k);       //first indicates the beginning of a common subsequence
          len++;
          flag=True;}
        **else if** (flag) {
          flag=False;
          *len*=0;}
        k++;
      }
      *diag-set=diag-set* ∪ {(first, len)};      //saving every sub-diagonal of continuous 1's in *diag-set*
    }
    **2-2)** Look for continuous 1's in diagonals below the main diagonal
    *len*=0;
    **for** (i=1 to |t|){
      k=i;
      flag=False;
      **for** (j=1 to |s|) {
        **if** ($M_{s,t}$[j,k]==1){
          **if** (!flag) first=(j,k)
          len++;
          flag=True;}

```
        else if (flag) {
            flag=False;
            len=0; }
        k++;
      }
    diag-set=diag-set ∪ {(first, len)} ;     // saving every sub-diagonal of continuous 1's in diag-set
  }
```

**3)** Find non-overlapping sub-diagonals of continuous 1's in *diag-set*
  Repeat steps 3.1 to 3.3 as far as there are no unmarked sub-diagonals
  **3-1) if** (some of the sub-diagonals $((i,j), len)$, have the same $len$)
        select the one that has the smallest $(i - j)$
      **else** select the longest one.
  **3-2)** Mark the selected *sub-diagonal*
  **3-3)** Modify the other sub-diagonals to discard overlaps with the selected *sub-diagonal*
      **while** (there exist any element in *diag-set*){
          ((i,j),len)=find_longest(*diag-set*)
          **for** each element $((i',j'), len')$ in diagonal-set{
              **if** $((i \leq i' \leq i + len - 1) \&\& (i \leq i' + len' - 1 \leq i + len - 1))$
                  remove $((i',j'), len')$ from diagonal-set
              **else if**$((j \leq j' \leq j + len - 1) \&\& (j \leq j' + len' - 1 \leq j + len - 1))$
                  remove $((i',j'), len')$ from diagonal-set
              **else if**$(i' + len' - 1 \leq i + len - 1)$
                  $len'=(i' + len' - 1) - i;$
              **else if**$(j' \leq j + len - 1)$
                  $len'=(j' + len' - 1) - (j + len - 1);$ }
          mark$((i,j), len)$ in *diag-set;*}

**4)** Compute the similarity between *s* and *t*.
  **for** *k* elements ((i,j),len) in *diag-set*
      $sim(s,t) += len;$
  $sim(s,t) = (sim(s,t))/(k \times max(|s|, |t|));$

The resultant SVM model uses the following relation to classify an execution path $e_j$:

$$f(e_j) = sign(\sum_{\rho=1}^{N_s} \alpha_\rho C_\rho k(x_\rho, e_j) - b) \tag{15}$$

In the above relation, $N_s$ indicates the number of support vectors for the program. $\alpha_\rho$ indicates the Lagrange multiplier for $\rho$th support vector. From this equation it is possible to see that the Lagrange multiplier $\alpha_\rho$ associated with the training point $x_\rho$ expresses the strength of $x_\rho$ in the final decision function. A remarkable property of this classifier is that it associates only a subset of the training points with a non-zero $\alpha_\rho$. These points are called 'support vectors' and are the points that lie closest to the separating hyperplanes. According to the sign of $f(e_j)$, SVM estimates the state of an execution path $e_j$.

## 4 Experiments

We have conducted some experiments to assess the performance of the proposed technique. Throughout the experiments, we used Tarantula [10], Ochiai [8], and Naish [9] methods to assess the proposed technique in enhancing fault localization. We have used the Siemens suite, gzip, grep, sed, space, and flex as our subject programs. The programs are downloaded from Software Infrastructure Repository (SIR). A brief description of the test suites is presented in Table 3.

Table 3. A brief description of subject programs

| Subject programs | Faulty versions | Lines | Test cases | Description |
|---|---|---|---|---|
| Siemens Suite | 132 | 2903 | 21631 | *lexical analyzer* |
| Gzip | 16 | 5159 | 217 | *compressor* |
| Grep | 18 | 9089 | 809 | *text processor* |
| Sed | 17 | 9298 | 370 | *text processor* |
| Flex | 6 | 13892 | 567 | *Lexical parser* |
| Space | 38 | 6199 | 13585 | *Array definition language* |

## 4.1 Evaluation Metrics

The proposed technique is evaluated on both prediction and faults localization's performance. In this regard, the following metrics are used in this paper:

1) '*The EXAM score*' gives the percentage of statements that need to be examined until the first faulty statement is reached.
2) 'Average number of statements examined' metric gives the average number of statements that need to be examined with respect to a faulty version to find the bug.
3) '*Wilcoxon signed-rank test*' metric. To conduct an evaluation based on sound statistics, we make use of the Wilcoxon signed-rank test, an alternative to the paired Student's t-test when a normal distribution of the population cannot be assumed [35]. Since we aim to show that our proposed method is more effective than other compared methods, we evaluate the one-tailed alternative hypothesis that the other techniques require the examination of an equal or greater number of statements than our method. The null hypothesis in this case specifies that the other techniques require the examination of a number of statements that is less than required by our method. Thus, the alternative hypothesis is that our method will require the examination of fewer statements than the other techniques to locate faults, implying that is more effective.
4) We also used two metrics that introduced by Masri et al. [19] to measure the effectiveness of fault localization techniques.
   a) Safety Change: Safety indicates the relative suspiciousness of the faulty code. Assuming that $f$ is the faulty statement, we use the $score(f)$ and $score'(f)$ to present the suspiciousness score of $f$ calculated with a fault localization technique before and after addressing coincidentally correct test cases, using our proposed approach. If $score'(f) > score(f)$, we consider that the safety of the fault localization technique is increased. Otherwise, the safety of the technique is reduced.
   b) Precision Change: Precision describes the reduction in the search space for the faulty statement. $R$ denotes the numbers of statements that has a larger suspiciousness score than the faulty statement. If $R$ is reducing after addressing coincidentally correct tests, using our proposed approach, the precision is considered to be improved. Intuitively, the change of precision can be observed by checking the rank of the faulty statement.
5) We also compute three major measurement metrics, i.e., precision, recall, and F-measure, to evaluate the performance of our proposed method.

Since for all techniques used in experiments, the same suspiciousness value may be assigned to multiple statements, the results are provided in two different levels of effectiveness – the ''best'' and the ''worst''. In all our experiments we assume that for the ''best'' effectiveness we examine the faulty statement first and for the ''worst'' effectiveness we examine the faulty statement last.

There are two strategies to deal with identified CC test cases. The first is to remove them and calculate the suspiciousness scores using the remaining test cases [19]. The second is to flip the outcome of identified CC test cases to '*failed*' when conducting fault localizations [19]. Since, removal of test cases causes loss of information by reducing the coverage of the test suite used for fault localization, we used the flipping strategy. The results demonstrate that when CC test cases are present, the faulty statements are ranked as less suspicious than when they are not present.

## 4.2 Other Techniques to Coincidental Correctness Detection Used in Comparisons

We make use of two other methods of coincidental correctness identification to compare with our proposed method. The first is a clustering-based and the second is an ensemble-SVM based method. In the following, these two methods are briefly introduced.

### 4.2.1 Clustering-based method

Miao et. al., [18], propose a clustering-based strategy to identify coincidental correctness. The key rationale behind this strategy is that test cases in the same cluster have similar behaviors [30, 31]. Therefore, a passed test case in a cluster, which contains failed test cases, is highly possible to be coincidental correctness because it has the potential to cover the faulty elements as those failed test cases do. Note that similar to our method, their approach is based on the single-fault assumption.

### 4.2.2 SVM-Ensemble

Xue et. al., [20], propose a technique to identify coincidentally correct test cases. The proposed technique combines support vector machines and ensemble learning to detect mislabeled cases, i.e. coincidentally correct test cases. They argue that the problem of identification of coincidentally correct test cases is, in fact, an instance of identification of mislabeled data in machine learning. In the scenario, they regard coincidentally correct test cases as mislabeled data, i.e. test cases whose test status should be flipped into "failed".

## 4.3 Experimental Results

We first evaluate the performance of fault localization methods on different subject programs without any dealing with coincidental corrects tests. For example, we can see the performance of Ochiai method in Table 4. We then attempt to identify the coincidentally correct tests using our proposed method and relabel them from 'passing' to 'failing'. Then, we apply fault localization methods on different subject programs again to investigate the amount of obtained improvement.

Table 4  Average number of statements examined (Best case)

|  | Siemens | Gzip | Grep | Sed | Space | Flex | Sum | Improvement |
|---|---|---|---|---|---|---|---|---|
| $Tarantula$ (before) | 38.41 | 109.36 | 306.82 | 244.36 | 114.56 | 94.21 | 907.72 | 14.56% |
| $Tarantula'$ (after) | 31.25 | 86.5 | 274.62 | 206.75 | 99.63 | 76.72 | 775.47 | |
| $Ochiai$ (before) | 26.57 | 89.10 | 194.50 | 121.45 | 84.35 | 73.21 | 589.18 | 17% |
| $Ochiai'$ (after) | 21.17 | 68.56 | 168.4 | 101.74 | 67.51 | 61.48 | 488.86 | |
| $O$ (before) | 21.41 | 78.16 | 151.06 | 84.45 | 74.12 | 66.24 | 475.44 | 18% |
| $O'$ (after) | 18.64 | 62.26 | 128.42 | 66.29 | 59.77 | 54.63 | 390.01 | |

Table 5  Average number of statements examined (Worst case)

|  | Siemens | Gzip | Grep | Sed | Space | Flex | Sum | Improvement |
|---|---|---|---|---|---|---|---|---|
| $Tarantula$ (before) | 64.22 | 202.14 | 412.14 | 368.92 | 148.26 | 162.2 | 1357.88 | 11.67% |
| $Tarantula'$ (after) | 53.40 | 182.56 | 376.24 | 326.62 | 121.78 | 138.72 | 1199.32 | |
| $Ochiai$ (before) | 53.62 | 176.90 | 294.06 | 198.05 | 104.89 | 114.33 | 941.85 | 14.67 % |
| $Ochiai'$ (after) | 41.66 | 147.46 | 262.56 | 166.58 | 89.85 | 98.52 | 803.63 | |
| $O$ (before) | 64.72 | 165.52 | 251.63 | 206.05 | 98.89 | 121.63 | 908.44 | 13.73% |
| $O'$ (after) | 50.42 | 144.8 | 226.78 | 184.26 | 74.67 | 102.78 | 783.71 | |

Tables 4 and 5 present the average number of statements that need to be examined by each fault localization technique across each of the subject programs for best and worst cases, respectively. For example, the average number of statements examined by Ochiai with respect to the all faulty versions of Gzip is 89.10 and 176.90 in best and worst cases, respectively. These values decrease to 68.56 and 147.46 when we deal with coincidentally correct tests using our proposed method. Since the data provided in Tables 4 and 5 represent the average of all faulty versions for each program, we also used the Wilcoxon signed-rank test to allow a better understanding how the two samples (without dealing the coincidentally correct tests; and with dealing) behave.

Table 6 presents data comparing our proposed method to the other techniques using the Wilcoxon signed-rank test. Each entry in the table gives the confidence with which the alternative hypothesis (that our proposed method requires the examination of fewer statements than a given technique to find faults, thereby making it more effective) can be accepted with respect to a selected program.

**Table 6** Confidence with which it can be claimed that our method is more effective than others

| Technique | Siemens | Gzip | Grep | Sed | Space | Flex |
|---|---|---|---|---|---|---|
| Tarantula | 97.50 % | 98.48 % | 97.96 % | 96.55 % | 96.95 % | 91.40 % |
| Tarantula – Clustering | 92.85 % | 95.20 % | 92.75 % | 93.25 % | 92.90 % | 89.74 % |
| Tarantula – SVM-Ensemble | 86.28 % | 87.95 % | 90.44 % | 89.35 % | 89.45 % | 85.60 % |
| Ochiai-Star | 96.88 % | 98.05 % | 97.66 % | 96.09 % | 96.50 % | 90.99 % |
| Ochiai - Clustering | 92.19 % | 94.66 % | 92.22 % | 92.88 % | 92.22 % | 87.60 % |
| Ochiai - SVM-Ensemble | 85.50 % | 87.27 % | 90.19 % | 88.50 % | 87.81 % | 84.56 % |
| O | 97.50 % | 98.48 % | 97.96 % | 96.55 % | 96.95 % | 91.40 % |
| O – Clustering | 92.85 % | 95.20 % | 92.75 % | 93.25 % | 92.90 % | 89.74 % |
| O – SVM-Ensemble | 86.28 % | 87.95 % | 90.44 % | 89.35 % | 89.45 % | 85.60 % |

To take an example, it can be said with 92.75% confidence that our method is more effective than Clustering-based method on the Grep program, when using Tarantula fault localization method. Similar observations can also be made for most of the scenarios in Table 6 with more than 90% confidence except for a few, such as being more effective than SVM-Ensemble with 87.27% confidence on the Gzip program, when using Ochiai method.

Figure 1 presents the safety and precision change after applying our proposed method on Ochiai fault localization. The x-coordinate is the subject program and the y-coordinate is the percentage of versions, which have a safety improvement or precision improvement.

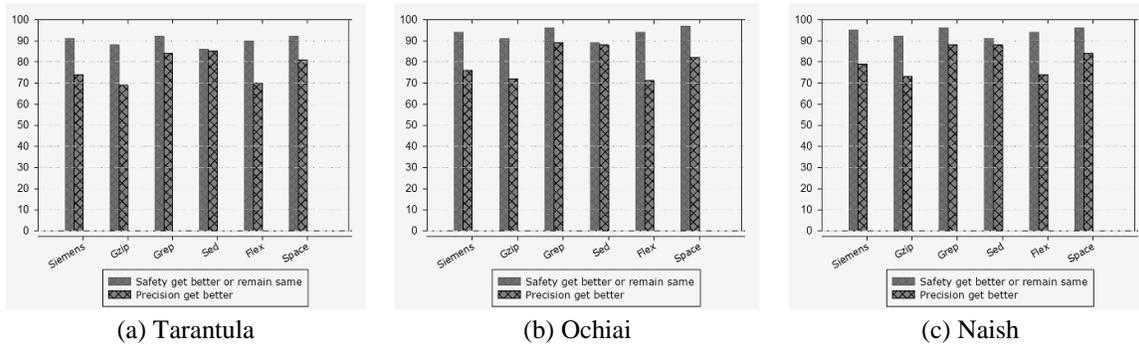

(a) Tarantula      (b) Ochiai      (c) Naish

Figure 1. The improvement of safety and precision measures

For the majority of programs, the safety is getting better or at least keeps a same value. It means that the suspiciousness scores of faulty statements are increasing. There are a few versions that the safety value decrease. However, the reductions are relatively small. For about 77% versions of programs, applying our proposed method results in a precision improvement, which means the rank of faulty statement is up. Although the precision of 23% of versions is decreased, the decrements are relatively small. In summary, the safety of the majority of programs is improved with only several versions staying the same or getting worse. There are 77% of all versions that have a precision improvement with an average rate of 7.3%, and 23% of versions are decreased with an average rate of 1.3 % which is relatively small. The results of safety and precision change indicate that our approach can improve the safety and precision of fault localization to some degree.

To evaluate the significance of improvement by our proposed method, we applied the paired Wilcoxon tests between our proposed and other two compared methods when using Ochiai fault localization method. In this regard, we carried out the one-tailed alternative hypothesis to verify that our method leads to more improvement in safety and precision.

The p-values of all tests between our method and Clustering-based method range from 0.1090 to 0.1405, and between our method and SVM-Ensemble-based method range from 0.1120 to 0.1678. Therefore, we can accept the hypothesis with confidence level 0.8595 of the test between our method and Clustering-based method and with confidence level 0.8322 of the test between our method and SVM-Ensemble-based.

Tables 7 and 8 present data comparing our proposed method to the other compared techniques using the Wilcoxon signed-rank test. Each entry in the tables gives the confidence with which the alternative hypothesis (that leads to more improvement in precision and safety than other technique, thereby making it more effective) can be accepted with respect to a selected program.

**Table 7** Confidence with which it can be claimed that slicing-based method is more effective than Others-Safety

| Technique | Siemens | Gzip | Grep | Sed | Space | C-Math |
|---|---|---|---|---|---|---|
| Ochiai | 94.21 % | 93.55 % | 91.26 % | 88.95 % | 90.44 % | 91.55 % |
| Ochiai - Clustering | 87.25 % | 89.05 % | 89.10 % | 85.95 % | 87.74 % | 88.80 % |
| Ochiai - SVM-Ensemble | 88.80 % | 84.90 % | 85.66 % | 83.22 % | 86.55 % | 85.24 % |

**Table 8** Confidence with which it can be claimed that slicing-based method is more effective than Others-Precision

| Technique | Siemens | Gzip | Grep | Sed | Space | C-Math |
|---|---|---|---|---|---|---|
| Ochiai | 96.50 % | 97.45 % | 93.55 % | 96.74 % | 92.14 % | 93.74 % |
| Ochiai - Clustering | 94.80 % | 93.88 % | 92.57 % | 91.55 % | 90.80 % | 91.55 % |
| Ochiai - SVM-Ensemble | 92.66 % | 91.55 % | 88.57 % | 89.55 % | 88.44 % | 89.80 % |

In another experiment, we attempt to evaluate the proposed and other compared methods using precision, recall, and F-measure to show how well each method works in identifying coincidentally correct tests. Suppose that $Ticc$ and $Tcc$ denote the set of test cases determined to be coincidentally correct and the set of actual coincidentally correct test cases, respectively. Precision, Recall and F-measure are measured by the expressions:

$$\text{Precision} = \frac{|Tcc \cap Ticc|}{|Ticc|} \quad, \quad Recall = \frac{|Tcc \cap Ticc|}{|Tcc|} \quad, \quad F-Measure = 2 \times \frac{\text{Precision} \times Recall}{\text{Precision} + Recall} \quad (16)$$

Table 9 reports the performance of our proposed and other compared methods in identifying coincidentally correct tests using Precision, Recall, and F-Measure metrics. Overall, the average precision (87%), the recall (89.16%), and the F-measure (88.05%) ratios are high. High precision indicates that for most cases, both the true coincidentally correct and true non-coincidentally correct tests are identified properly. The recall ratio is 89.16% pointing out that we can effectively identify most of the true coincidentally correct tests. A method with a low recall value means that only a subset of all coincidentally correct test cases was classified correctly. As a result, non-faulty statements executed by the classified subset of the coincidentally correct test cases obtain a suspiciousness score equal or higher to that of the faulty statement. Thus, a lower recall results in lower effectiveness.

**Table 9** Performance of the proposed and other methods w.r.t Precision, Recall and F-Measure

| Program | Our proposed approach | | | Clustering-based Approach | | | SVM-Ensemble Approach | | |
|---|---|---|---|---|---|---|---|---|---|
| | Precision % | Recall % | F-Measure % | Precision % | Recall % | F-Measure % | Precision % | Recall % | F-Measure % |
| Siemens | 87 | 91 | 88.95 | 77 | 72 | 74.81 | 80 | 82 | 80.98 |
| Gzip | 93 | 90 | 91.47 | 78 | 68 | 73.55 | 82 | 79 | 80.24 |
| Grep | 88 | 89 | 88.49 | 79 | 65 | 72.28 | 79 | 76 | 77.24 |
| Sed | 86 | 92 | 88.90 | 82 | 74 | 78.55 | 86 | 83 | 74.98 |
| Space | 85 | 89 | 87.00 | 76 | 70 | 73.21 | 81 | 77 | 78.90 |
| Flex | 83 | 84 | 83.49 | 73 | 64 | 68.90 | 78 | 77 | 77.35 |
| Average | 87 | 89.16 | 88.05 | 77.50 | 68.83 | 73.60 | 81 | 79 | 78.28 |

To show that our proposed method is more effective in identifying coincidentally correct tests than the other compared methods, we also carried out the one-tailed alternative hypothesis to verify that our method leads to greater values in precision, recall and F-measure employing the Ochiai method. The p-values of all tests between our method and Clustering-based method range from 0.119 to 0.138, and between our method and SVM-Ensemble-based method range from 0.135 to 0.254. Therefore, we can accept the hypothesis with confidence level 0.862 of the test between our method and Clustering-based method and with confidence level 0.746 of the test between our method and SVM-Ensemble-based. We also present the evaluation of proposed method and two other compared works with respect to the *EXAM* score. Figure 2 represents the *EXAM* score of Ochiai [8] method on subject programs, after addressing CC tests using our proposed, SVM-Ensemble[20] and Clustering-based method [18].

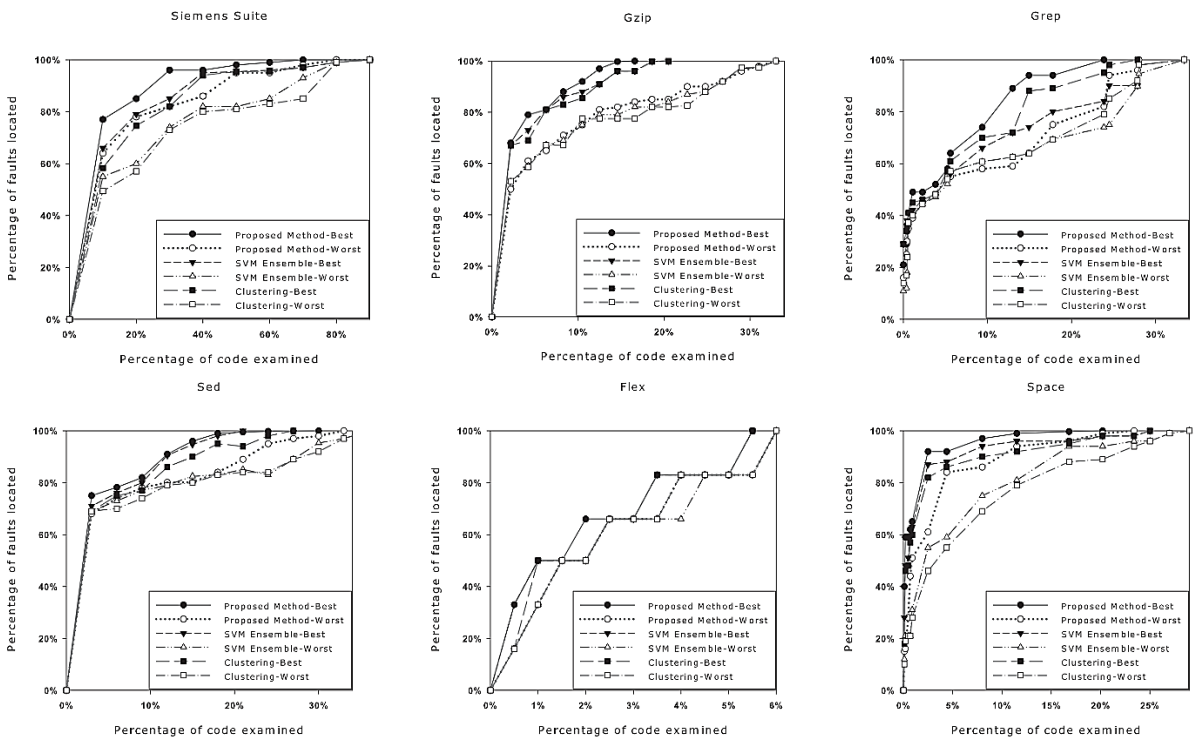

Figure 2. EXAM score-based comparison of techniques using Ochiai Method

## 5 Related Works

The study of CC began with the analysis of faults status [26], and it was defined when a fault is executed but no failure is detected. Masri [19] introduced three strategies of cleansing test suites from CC to enhance fault localization. In the first strategy, they first identify the set of program elements that are likely to be correlated with coincidentally correct tests; an element in this set is called $cc_e$. Then, any test that induces one or more $cc_e$, called a $cc_t$, will be considered a potential CC test. The second strategy assumes that a $cc_t$ containing a large number of $cc_e$'s and/or $cc_e$'s with a high average suspiciousness is more likely to be a CC test than another that does not. The third strategy partitions the $cc_t$'s into two clusters based on the similarity of the suspicious $cc_e$'s they execute, then ignores one of them based on further criteria. All three strategies suffered from a high rate of false positives, while the third strategy was the better performer. More recently, Masri et al. [27] presented a multivariate visualization-based technique that enables the user to identify CC tests. Furthermore, Miao [18] employed a k-means clustering-based technique to identify CC. A coverage refinement approach is presented in Wang et al. [13] to reduce the influence of CC on fault localization. The work introduces a concept called context pattern, which is unique for each fault type and describes the program behavior before and after the faulty code. In [20], Xue et al. proposed a technique which first employs support vector machine ensemble to detect coincidental correctness, then trimming the test suit by removing or flipping the detected coincidentally correct test cases. One problem with these ensemble-based methods is that the underlying models are built from a training set that contains mislabeled examples. Another problem is the requirement that each base classifier must be independent and has type 1 and type 2 error rates of less than 50% [29], which may not hold when the level of noise is sufficiently high.

# 6 Threats to Validity

Threats to external validity arise when the results of the experiment are unable to be generalized to other situations. While it is true that the evaluation of the proposed method is based on empirical data and therefore we may not be able to generalize our results to all programs; it is for this reason that we observed the effectiveness of the methods across such a broad spectrum of programs. Each of the subject programs varies greatly from the other on size, function, number of faulty versions studied, etc. The threat to construct validity is our measurements for the experiment. To minimize the threat, we introduce widely used measurements in fault localization. For the evaluation of prediction, we used accuracy, precision, and recall to measure the performance. However, in practice, there may be other metrics and representation demonstrating how well a classifier performs. We use the Tarantula, Ochiai, and Naish methods to compute the suspiciousness scores for each statement. In practice, there might be other computation methodologies to assess the suspiciousness of statements in a given program.

# 7 Conclusion

In this paper, we present a kernel-based approach for detecting CC runs in a test suite. To build a kernel function for an SVM classifier, a new sequence-matching algorithm is applied to find the common non-overlapping sub-paths among different program execution paths. Then, the trained SVM classifier is used to address the identification of CC test cases through comparing program execution paths with the hyperplane. The main problem is the high similarity between program executions that may lead to nonlinear separability of the corresponding vectors. To resolve this difficulty, SVM applies a kernel function to increase the dimensionality of space such that the execution vectors within the space could be linearly separated. In the proposed kernel function, non-overlapping common subsequences of the maximum length between the execution paths are considered as new dimensions of the feature space. In this feature space, the similarities between execution paths could be easily measured regarding the number of common subsequences. These similarities are a fundamental basis to construct the SVM classifier. The results of experiments show that the proposed technique is highly effective in addressing the CC problem and reducing the effort programmers spent on fault localization.